\documentclass[
  aps,prd,reprint,10pt,
  superscriptaddress,
  nofootinbib
]{revtex4-2}

\usepackage{amsmath,amssymb}
\usepackage{graphicx,array}
\usepackage{fontawesome}
\usepackage{silence}

\usepackage[T1]{fontenc}
\usepackage{newtxtext,newtxmath}
\usepackage[cal=cm]{mathalfa}
\usepackage{bm}

\usepackage[dvipsnames]{xcolor}
\usepackage[colorlinks=true,allcolors=Blue]{hyperref}

\usepackage{orcidlink}

\begin{document}
\title{Sensitivity of calibration-free DESI DR2 BAO--supernova constraints to a distance-duality tilt}

\author{Mikel Artola\orcidlink{0009-0007-9068-1995}}
\email[Contact author: ]{mikel.artola@ehu.eus}
\affiliation{Department of Physics, \href{https://ror.org/000xsnr85}{University of the Basque Country EHU}, 48940 Leioa, Spain}
\affiliation{EHU Quantum Center, \href{https://ror.org/000xsnr85}{University of the Basque Country EHU}, 48940 Leioa, Spain}

\author{St\'ephane Fay\orcidlink{0000-0002-6601-4304}}
\email[Contact author: ]{steph.fay@gmail.com}
\affiliation{Astronomy Department, \href{https://ror.org/00maat333}{Palais de la D\'ecouverte}, Avenue Franklin Roosevelt, 75008 Paris, France}

\author{Leandros Perivolaropoulos\orcidlink{0000-0001-9330-2371}}
\email[Contact author: ]{leandros@uoi.gr}
\affiliation{Department of Physics, \href{https://ror.org/01qg3j183}{University of Ioannina}, GR-45110, Ioannina, Greece}

\begin{abstract}
Recent measurements from DESI baryon acoustic oscillations (BAO) have reported a moderate preference for dynamical dark energy, challenging the $\Lambda$CDM paradigm. Nonetheless, such a trend could reflect a tension between the shape and the offset of the Hubble diagrams predicted by current BAO and Type Ia supernovae (SN) data, which may underpin a possible breakdown of the cosmic distance duality relation linking luminosity distances $D_{\rm L}$ and angular-diameter distances $D_{\rm A}$. We explore this possibility by parametrizing deviations from this relation as $\eta(z)\equiv D_{\rm L}/[{(1+z)}^2D_{\rm A}]=\alpha{(1+z)}^\varepsilon$, and investigate how the preference for a Chevallier-Polarski-Linder (CPL) dynamical dark energy is modified once $\varepsilon\neq0$ is allowed. Using DESI DR2 and uncalibrated SN data (Pantheon$+$, DES-Dovekie, Union3, and Union3.1), we find that allowing $\varepsilon$ to vary freely in both $\Lambda$CDM and CPL reduces the CPL preference with respect to $\Lambda$CDM by a larger amount than the addition of spatial curvature $\Omega_k$; only Union3.1 exhibits comparable reductions. We also find that the preference reduction is robust to alternative $\eta(z)$ parametrizations, and that the $\eta(z)$ constraints remain stable under the removal of BAO tracers or low-redshift cutoffs for Pantheon$+$, whereas the DES-Dovekie preference largely disappears once $z<0.1$ SN are removed. In a separate calibrated analysis, illustrative Planck sound-horizon and SH0ES absolute-magnitude priors give $\alpha\simeq0.942\pm0.014$, quantifying the normalization mismatch under those assumptions. We also discuss photon-loss and varying-$G$ interpretations: ordinary positive opacity cannot explain the negative central tilts $\varepsilon$ found for a $\Lambda$CDM background, while local gravity tests strongly constrain the illustrative unscreened varying-$G$ scenario.
\end{abstract}
\maketitle

\section{\label{sec:introduction}Introduction}

Since the discovery of the accelerated expansion of the Universe by Type Ia supernovae (SN)~\cite{SupernovaSearchTeam:1998fmf, SupernovaCosmologyProject:1998vns}, the cosmological constant ($\Lambda$) together with cold dark matter (CDM), namely the $\Lambda$CDM model, has been the standard paradigm to describe the evolution and formation of large scale structures. Its success has been consistently reaffirmed by multiple observational probes, including baryon acoustic oscillations (BAO)~\cite{SDSS:2005xqv, 2dFGRS:2005yhx} and cosmic microwave background (CMB)~\cite{Boomerang:2000efg, WMAP:2003elm} observations. However, with the arrival of increasingly high precision data, the $\Lambda{\rm CDM}$ model has started to reveal its struggles in simultaneously explaining all observations.

Perhaps the most critical challenge is the Hubble tension~\cite{Planck:2018vyg, Riess:2021jrx}, that is, the severe discrepancy between early- and late-time values predicted for the Hubble constant $H_0$. CMB data by Planck, South Pole Telescope and the Atacama Cosmology Telescope~\cite{SPT-3G:2025bzu} find $H_0 = 67.24 \pm 0.35\, {\rm km}\, {\rm s}^{-1}\, {\rm Mpc}^{-1}$, whereas the H0DN collaboration and their Local Distance Network~\cite{H0DN:2025lyy} find $H_0 = 73.50 \pm 0.81\, {\rm km}\, {\rm s}^{-1}\, {\rm Mpc}^{-1}$, translating into a $7.1\sigma$ statistical discrepancy between the two measurements. Although numerous theoretical models have been proposed to alleviate this and many other tensions (see e.g.~\cite{CosmoVerseNetwork:2025alb} for an exhaustive review), no conclusive candidate has been found yet~\cite{Vagnozzi:2023nrq, Poulin:2024ken}.

Furthermore, recent BAO measurements from the Dark Energy Spectroscopic Instrument (DESI)~\cite{DESI:2024mwx, DESI:2025zgx}, combined with CMB and various SN data (Pantheon$+$~\cite{Scolnic:2021amr, Brout:2022vxf}, Union3~\cite{Rubin:2023jdq}, and DES-SN5YR~\cite{DES:2024jxu}) have shown a preference for dynamical dark energy, ranging from $2.8$--$4.2\sigma$ in their Data Release 2 (DR2)~\cite{DESI:2025zgx} depending on the supernovae sample. The strongest preference was reported by the DES-SN5YR combination, though it has been settled at around $3\sigma$ after the recent DES-Dovekie reanalysis~\cite{DES:2025sig}.%
\footnote{During the elaboration of this manuscript, the Supernovae Unite catalog has presented its analysis~\cite{Camilleri:2026rer,Lee:2026dum}, combining the Pantheon$+$ and DES-Dovekie samples in an internally consistent manner. Although at the current time no public data is available, the collaboration reported a $3.1$--$3.3\sigma$ preference (depending on whether $\Delta \chi_{\rm min}^2$ or $\Delta \chi_{\rm MAP}^2$ is used) for the Chevallier-Polarski-Linder~\cite{Chevallier:2000qy, Linder:2002et} dynamical dark energy compared with $\Lambda$CDM.}
Removing CMB from the analysis places this preference at $1.7$--$3.3\sigma$~\cite{DESI:2025zgx}. At present, there is no clear consensus about whether this could be a genuine signal: some Bayesian analyses report no preference over $\Lambda$CDM~\cite{Ong:2025utx,Ong:2026tta,LozanoTorres:2026xxx,Zapata:2026evt}, while others report a mild preference~\cite{DESI:2025wyn}.

One possibility is that the apparent preference for dynamical dark energy is caused by internal inconsistencies among the datasets combined in the inference~\cite{Bousis:2024rnb, Colgain:2024mtg, Wang:2025bkk, Afroz:2025iwo, Sapone:2026iwc}. From a more fundamental perspective, this could be signaling a discrepancy between two kinds of distance measurements: those inferring the angular diameter distance $D_{\rm A}(z)$, obtained from the apparent angular size of an object with known physical size, and those inferring the luminosity distance $D_{\rm L}(z)$ from the flux of known intrinsic brightness sources. This comparison is usually performed under the assumption that the cosmic distance duality relation (CDDR) holds at all redshifts: $D_{\rm L}(z) = {(1 + z)}^2 D_{\rm A}(z)$. This relation is valid in (i) metric theories of gravity where (ii) photons travel along unique geodesics and (iii) their number is conserved. The first two conditions establish the well-known Etherington's reciprocity theorem~\cite{Etherington:1933asu} relating the source and observer area distances as $D_{\rm S}(z) = (1+z) D_{\rm A}(z)$; the additional photon-number conservation condition further gives $D_{\rm L}(z) = (1+z) D_{\rm S}(z)$. If any of these three assumptions does not hold, the CDDR need not be satisfied. From a physical perspective, CDDR violations can appear in: non-pseudo-Riemannian manifolds~\cite{Santana:2017zvy, Azevedo:2021npm} (e.g.~torsion-based theories~\cite{Cai:2015emx} or in energy momentum squared gravity~\cite{Cipriano:2024jng}); scenarios beyond the Standard Model with photons coupled to additional particles (e.g.~axion-like particles~\cite{Csaki:2001yk, Bassett:2003vu, Bassett:2003zw, Avgoustidis:2010ju, Dhawan:2025mer} or gravitons in external magnetic fields~\cite{Raffelt:1987im, Chen:1994ch, Cillis:1996qy}); chameleon fields coupled to matter~\cite{Khoury:2003aq, Khoury:2003rn}; photon--chameleon mixing~\cite{Burrage:2007ew}; Kaluza-Klein models~\cite{Deffayet:2000pr}; or even scenarios in which variations of fundamental constants occur~\cite{Barrow:1999is, Goncalves:2019xtc, Lee:2021xwh, Santos:2025gjf}, among others~\cite{Avgoustidis:2011aa, Ellis:2013cu, Giesel:2022org}. More prosaic explanations could involve photon scattering off free electrons and dust in the intergalactic medium, intrinsic source evolution, or calibration effects.

In this work we investigate the CDDR violation by considering a power-law parametrization. Our aim is to perform a sensitivity study by comparing how much the preference for Chevallier-Polarski-Linder (CPL)~\cite{Chevallier:2000qy,Linder:2002et} dynamical dark energy over $\Lambda$CDM is reduced when violations of CDDR are allowed in both models simultaneously. We focus on data combinations involving only BAO from the DESI DR2~\cite{DESI:2025zgx} and four state-of-the-art SN samples separately (Pantheon$+$~\cite{Scolnic:2021amr, Brout:2022vxf}, DES-Dovekie~\cite{DES:2025sig}, Union3~\cite{Rubin:2023jdq}, and Union3.1~\cite{Rubin:2026qdt,Hoyt:2026fve}). The main analysis uses no external calibration of the Hubble constant $H_0$, sound horizon $r_{\rm d}$, or supernovae absolute magnitude $M_{\rm B}$. Our contribution relies on this calibration-free comparison across all four SN samples, together with controlled tests of alternative parametrizations, spatial curvature, BAO tracer selection, and SN redshift range, and complements the calibrated distance-duality analysis of Refs.~\cite{Teixeira:2025czm, Dhawan:2025mer} and earlier tests~\cite{Tonghua:2023hdz, Wang:2024rxm, Yang:2025qdg, Kanodia:2025jqh, Sapone:2026iwc, Martinelli:2026wjp}.

The structure of the manuscript is as follows. Sec.~\ref{sec:background_and_datasets} defines the modified distance-duality relation, the background cosmologies, the datasets, and the statistical comparison. Sec.~\ref{sec:BAO_SN_results} presents the calibration-free results and sensitivity tests. Sec.~\ref{sec:Hubble_tension} adds cosmic chronometers and then considers a separate diagnostic with Planck and SH0ES calibration priors. Sec.~\ref{sec:physics} discusses illustrative physical realizations of CDDR violation, and Sec.~\ref{sec:conclusions} summarizes the findings and limitations.

\section{\label{sec:background_and_datasets}Theoretical background}

\subsection{\label{sec:distance_duality}Cosmic distance duality relation}

The CDDR is a fundamental result relating the angular diameter distance $D_{\rm A}$ and the luminosity distance $D_{\rm L}$ at a redshift $z$. For any given metric theory of gravity in which photons follow unique null geodesics and their number is conserved, the CDDR must be satisfied:
\begin{equation}
    D_{\rm L}(z) = {(1+z)}^2 D_{\rm A}(z).
    \label{eq:etherington}
\end{equation}
This relation is independent of the specific background expansion history. Nonetheless, if the datasets considered in an observational analysis have unaccounted-for systematics (e.g.~evolution in the properties of supernova progenitors, intergalactic dust attenuation, or perhaps exotic physics), this could lead to an apparent violation of CDDR. To account for this possibility, we consider an effective distance-duality relation parametrized as
\begin{equation}
    \eta(z) \equiv
    \frac{D_{\rm L}(z)}{{(1 + z)}^2 D_{\rm A}(z)} =
    \alpha{(1 + z)}^\varepsilon,
    \label{eq:eta}
\end{equation}
where $\alpha$ controls the overall normalization and $\varepsilon$ parametrizes a redshift-dependent tilt. The power-law dependence may naturally arise in scenarios where photon numbers are not conserved; the normalization $\alpha$, on the other hand, should be understood as a mathematically equivalent way of capturing calibration discrepancies between data, rather than capturing CDDR violations with a physical origin. The exact CDDR is recovered upon setting $(\alpha, \varepsilon) = (1, 0)$, i.e.~$\eta(z)=1$.

\subsection{\label{sec:background} Background cosmologies}

We work within General Relativity, assuming a homogeneous, isotropic, and spatially flat Friedmann-Lemaître-Robertson-Walker (FLRW) background, unless otherwise specified. We consider contributions from (baryonic plus cold dark) matter and dark energy, with equation-of-state parameter $w = -1$ for $\Lambda{\rm CDM}$, and
\begin{equation}
    w(z) = w_0 + w_a \frac{z}{1+z}
\end{equation}
for a CPL dark energy component~\cite{Chevallier:2000qy,Linder:2002et}. The $\Lambda$CDM model is recovered in the limit $(w_0, w_a) = (-1, 0)$. The first Friedmann equation, which for a CPL dark energy ($w_0w_a{\rm CDM}$ hereafter) reads
\begin{equation}
    E^2(z) =
    \Omega_{\rm m} {(1\!+\!z)}^3 + (1 - \Omega_{\rm m}){(1\!+\!z)}^{3(1 + w_0 + w_a)} \exp\left(-\frac{3w_a z}{1+z}\right)
\end{equation}
fully determines the background evolution. Here $E(z) \equiv H/H_0$ is the dimensionless Hubble factor, with $H(z)$ the Hubble parameter and $H_0 \equiv H(z=0)$ the Hubble constant, and $\Omega_{\rm m}$ is the present-day fractional matter density.

Throughout we retain the standard FLRW geometrical relations used to model the BAO observables. In particular,
\begin{equation}
    D_{\rm M}(z) =
    (1+z) D_{\rm A}(z),
    \label{eq:transverse_and_angular_distances}
\end{equation}
and, for a spatially flat Universe,
\begin{equation}
    D_{\rm M}(z) =
    \frac{c}{H_0} I(z),
    \quad
    I(z) =
    \int_0^z \frac{{\rm d}z^\prime}{E(z^\prime)}.
\end{equation}
Therefore, the effective CDDR relation~\eqref{eq:eta} can be written as
\begin{equation}
    D_{\rm L}(z) =
    (1+z) \eta(z) D_{\rm M}(z).
    \label{eq:luminosity_distance}
\end{equation}
Since BAO from DESI provide transverse, Hubble, and volume-averaged distances (see Sec.~\ref{sec:datasets} below), the Hubble diagram these measurements constrain is independent of the CDDR violation.%
\footnote{More precisely, departures from the standard CDDR can be separated into two contributions. Following Ref.~\cite{Tiwari:2026pzk}, one may parametrize a violation of Etherington's reciprocity relation as $D_{\rm S}(z) = \eta_{\rm rec}(z) (1+z) D_{\rm A}(z)$, and photon-number non-conservation as $D_{\rm L}(z) = \eta_\gamma(z) (1+z) D_{\rm S}(z)$, so that $\eta(z) = \eta_{\rm rec}(z) \eta_\gamma(z)$. Our analysis does not attempt to distinguish these contributions, and therefore we omitted the details in the main text.}
However, distances to SN are determined through $D_{\rm L}$, and therefore, when combining both datasets, there appears a discrepancy in their predicted Hubble diagram modulated precisely by the deviation of $\eta(z)$ from unity.

\subsection{\label{sec:datasets}Datasets}

We now describe the datasets employed, as well as their corresponding likelihoods. Throughout, we assume the observational data are well described by Gaussian distributions~\cite{Dainotti:2023bwq}.

\subsubsection{\label{sec:bao}Baryon acoustic oscillations}

We employ the DESI DR2 data~\cite{DESI:2025zgx}, which includes $13$ measurements of the transverse comoving distance $D_{\rm M}$, the Hubble distance $D_{\rm H} \equiv c/H$, and the volume-averaged distance $D_{\rm V} \equiv (z D_{\rm M}^2 D_{\rm H})^{1/3}$, all normalized by the comoving sound horizon at the drag epoch $r_{\rm d}$, in the redshift range $0.295 \leq z \leq 2.33$. In the absence of an external calibration, BAO can only constrain the $H_0 r_{\rm d}$ combination, which we sample as a free parameter unless $H_0$ or $r_{\rm d}$ are provided or constrained by additional external datasets. The chi-squared statistic corresponding to BAO can be constructed as follows:
\begin{equation}
    \chi_{\rm BAO}^2 =
    {(\boldsymbol{D}_{\rm obs} - \boldsymbol{D}_{\rm th})}^T \cdot \boldsymbol{C}_{\rm BAO}^{-1} \cdot (\boldsymbol{D}_{\rm obs} - \boldsymbol{D}_{\rm th}).
\end{equation}
Here $\boldsymbol{D}_{\rm obs}$ is the vector of observed data points ($D_{\rm V}/r_{\rm d}$, or $D_{\rm M}/r_{\rm d}$ and $D_{\rm H}/r_{\rm d}$, depending on the observed redshift), $\boldsymbol{D}_{\rm th}$ is the theoretical prediction for the distances, and $\boldsymbol{C}_{\rm BAO}$ is the full covariance matrix as provided by the DESI collaboration.%
\footnote{\url{https://github.com/CobayaSampler/bao_data/tree/master/desi_bao_dr2}}

\subsubsection{\label{sec:sn}Type Ia supernovae}

We implement and independently consider four SN catalogs in our analysis: Pantheon$+$~\cite{Scolnic:2021amr, Brout:2022vxf}, DES-Dovekie~\cite{DES:2025sig}, Union3~\cite{Rubin:2023jdq}, and Union3.1~\cite{Rubin:2026qdt, Hoyt:2026fve}. Table~\ref{tab:data} contains the number of data points of each survey, together with the redshift range covered. For Pantheon$+$,%
\footnote{\url{https://github.com/PantheonPlusSH0ES/DataRelease/tree/main/Pantheon\%2B_Data/4_DISTANCES_AND_COVAR}.}
we exclude SN in Cepheid hosts and impose a Hubble diagram redshift cutoff $z_{\rm HD}>0.01$ to reduce the impact of peculiar velocities~\cite{Peterson:2021hel, Carr:2021lcj}; for the DES-Dovekie%
\footnote{\url{https://github.com/des-science/DES-SN5YR/tree/main/4_DISTANCES_COVMAT}.}
and the Union%
\footnote{\url{https://github.com/rubind/union3_release}.}
samples, we use the datasets as released.

\begin{table}[t]
\caption{Supernova likelihoods used separately with the 13 DESI DR2 BAO observables. Counts refer to retained distance entries (Pantheon$+$ and DES-Dovekie) or compressed nodes (Union, for which the {\scriptsize UNITY} pipelines were used). For Pantheon$+$ we impose a $z > 0.01$ low-redshift cutoff to reduce the impact of peculiar velocities, and remove the SN within Cepheid hosts. The remaining datasets are employed as released.}\label{tab:data}
\begin{ruledtabular}
\begin{tabular}{lcc}
Sample & Entries/nodes & Redshift range \\
\hline
Pantheon$+$ & 1580 & $0.010$--$2.261$ \\
DES-Dovekie & 1820 & $0.025$--$1.144$ \\
Union3 ({\footnotesize UNITY} 1.5) & 22 & $0.050$--$2.262$ \\
Union3.1 ({\footnotesize UNITY} 1.8) & 22 & $0.050$--$2.262$ \\
\end{tabular}
\end{ruledtabular}
\end{table}

The SN data vector contains corrected apparent magnitudes $m$ for the Pantheon$+$ sample, whereas for DES-Dovekie and the two Union releases the distance moduli $\mu$ are provided. The distance modulus is related to the luminosity distance through
\begin{equation}
    \mu(z) =
    25 + 5\log_{10} \left( \frac{D_{\rm L}(z)}{{\rm Mpc}} \right),
    \label{eq:distance_modulus}
\end{equation}
whereas the apparent magnitude and the distance modulus are related by $m = \mu + M_{\rm B}$, where $M_{\rm B}$ is the absolute magnitude of SN.\@ Since Pantheon$+$ and DES-Dovekie provide both the heliocentric redshifts $z_{\rm HEL}$ and the Hubble diagram redshifts $z_{\rm HD}$, we associate the former with the $(1+z)$ prefactor in Eq.~\eqref{eq:luminosity_distance}, and the latter with the $\eta(z)$ and the transverse comoving distance $D_{\rm M}(z)$. Then, the theoretical luminosity distance for the $i$th SN in a sample for spatially flat backgrounds becomes
\begin{equation}
    D_{{\rm L},i} =
    \frac{\alpha c}{H_0}(1+z_{{\rm HEL},i})(1+z_{{\rm HD},i})^\varepsilon I(z_{{\rm HD},i}),
\end{equation}
and the theoretical prediction vector $\boldsymbol{y}_{\rm th}$ for the $i$th entry can be written as
\begin{equation}
    y_{{\rm th},i} =
    5\log_{10}[(1+z_{{\rm HEL},i}) I(z_{{\rm HD},i})] + \varepsilon x_i + \nu,
    \label{eq:apparent_magnitude_or_distance_modulus}
\end{equation}
where $x_i = 5\log_{10}(1 + z_{{\rm HD},i})$, and $\nu$ is a normalization factor defined as
\begin{equation}
    \nu =
    25 + 5\log_{10} \left( \frac{c/H_0}{\rm Mpc} \right) + 5\log_{10}\alpha + \mathcal{Z}_{\rm SN}.
    \label{eq:nu}
\end{equation}
Here, $\mathcal{Z}_{\rm SN} = M_{\rm B}$ when comparing with apparent magnitudes, and is zero for distance moduli. In the absence of external calibrations of $H_0$ or $M_{\rm B}$, the individual parameters entering Eq.~\eqref{eq:nu} (including $\alpha$) cannot be determined, and SN instead constrain the total offset $\nu$. We retain this offset as a sampled parameter for simplicity, but do not report its values. Note that, in the uncalibrated setting, values compatible with $\varepsilon = 0$ do not necessarily imply $\alpha = 1$.

Then, the chi-squared statistic for the SN is given by
\begin{equation}
    \chi_{\rm SN}^2 =
    {(\boldsymbol{y}_{\rm obs} - \boldsymbol{y}_{\rm th})}^T \cdot \boldsymbol{C}_{\rm SN}^{-1} \cdot (\boldsymbol{y}_{\rm obs} - \boldsymbol{y}_{\rm th}),
\end{equation}
where $\boldsymbol{y}_{\rm obs}$ contains apparent magnitudes for Pantheon$+$ and distance moduli for the other compilations, $\boldsymbol{y}_{\rm th}$ is the corresponding vector of theoretical predictions, and $\boldsymbol{C}_{\rm SN}$ is the total (statistical plus systematic) covariance matrix of each sample.

\subsubsection{\label{sec:CC}Cosmic chronometers}

In Sec.~\ref{sec:Hubble_tension} we will also include cosmic chronometers (CC) in the analysis, for which we employ the expansion rate $H(z)$ measurements from Ref.~\cite{Moresco:2022phi} and construct the correlated stellar-population covariance following the steps exemplified in Ref.~\cite{Moresco:2020fbm}. The corresponding chi-squared statistic can be written as
\begin{equation}
    \chi_{\rm CC}^2 =
    {(\boldsymbol{H}_{\rm obs} - \boldsymbol{H}_{\rm th})}^T \cdot \boldsymbol{C}_{\rm CC}^{-1} \cdot (\boldsymbol{H}_{\rm obs} - \boldsymbol{H}_{\rm th}),
\end{equation}
with $\boldsymbol{H}_{\rm obs}$ the vector of observed data, $\boldsymbol{H}_{\rm th}$ the vector of predicted expansion rates, and $\boldsymbol{C}_{\rm CC}$ the covariance matrix. These measurements constrain the Hubble parameter $H_0$, allowing us to break the $H_0$--$r_{\rm d}$ degeneracy in BAO data. When both datasets are combined, we allow both $H_0$ and $r_{\rm d}$ to vary freely.

\subsection{\label{sec:model_comparison}Model comparison and sampling}

There are four cosmological models that we are interested in comparing: $\Lambda{\rm CDM}$, $w_0w_a{\rm CDM}$, $\Lambda{\rm CDM} + \varepsilon$, and $w_0w_a{\rm CDM}+\varepsilon$. Specifically, we want to quantify the improvement in the fit of $w_0w_a{\rm CDM}$ with respect to $\Lambda$CDM before and after allowing $\varepsilon$ to vary freely. To do so, we define and employ the following quantities throughout:
\begin{align}
    \Delta \chi_{\rm DE}^2 &=
    \min(\chi_{\Lambda{\rm CDM}}^2) - \min(\chi_{w_0w_a{\rm CDM}}^2), \label{eq:delta_chi2_de}\\
    \Delta \chi_{{\rm DE}\vert \varepsilon}^2 &=
    \min(\chi_{\Lambda{\rm CDM}+\varepsilon}^2) - \min(\chi_{w_0w_a{\rm CDM}+\varepsilon}^2), \label{eq:delta_chi2_de_eps}\\[.25em]
    f_{\rm red} &=
    1 - \Delta \chi_{{\rm DE}\vert \varepsilon}^2 / \Delta \chi_{\rm DE}^2. \label{eq:reduction_fraction}
\end{align}
The reduction fraction $f_{\rm red}$ is a descriptive change in the likelihood improvement from freeing the tilt $\varepsilon$ in both $\Lambda$CDM and $w_0w_a{\rm CDM}$ simultaneously. We also employ the Akaike information criterion (${\rm AIC}$)~\cite{Akaike:1974vps},
\begin{equation}
    {\rm AIC} = \chi_{\min}^2 + 2k,
\end{equation}
where $k$ includes all free cosmological and nuisance parameters, and follow the same definitions as in Eqs.~\eqref{eq:delta_chi2_de} and~\eqref{eq:delta_chi2_de_eps} when referring to $\Delta {\rm AIC}_{\rm DE}$ and $\Delta {\rm AIC}_{{\rm DE}\vert \varepsilon}$. Positive values of the statistics indicate preference in favor of $w_0w_a{\rm CDM}$ over $\Lambda{\rm CDM}$, and negative values indicate the reverse.
We opt for not reporting ${\rm AIC}$ preferences based on Jeffreys' scale~\cite{Jeffreys61, Trotta:2008qt}, which was designed for Bayes factors and whose thresholds can lead to biased conclusions~\cite{Nesseris:2012cq}. One possibility is to interpret $\Delta {\rm AIC}$ in terms of the Akaike weights~\cite{burnham2002model}, which for two models read $\omega = [1 + \exp(-\vert \Delta {\rm AIC} \vert/2)]^{-1}$ and give the probability that the preferred model is the best Kullback-Leibler model.

For orientation, we also convert an improvement $\Delta \chi^2$ into number of standard deviations $n_\sigma$ for the two nested models ($\Lambda{\rm CDM}$ and $w_0w_a{\rm CDM}$, and their extension with $\varepsilon$), as long as the regularity and asymptotic assumptions of Wilks' theorem~\cite{Algeri:2019lah} hold. The relative difference $\Delta \chi^2$ is approximately described by a $\chi^2$ distribution with $\Delta k$ degrees of freedom; in our case, for both $\Delta \chi_{\rm DE}^2$ and $\Delta \chi_{{\rm DE}\vert \varepsilon}^2$, this difference is $\Delta k=2$. Then, under the null hypothesis ($\Lambda{\rm CDM}$ or $\Lambda{\rm CDM}+\varepsilon$, respectively), and subject to these regularity assumptions, the probability of an improvement at least as large as the observed $\Delta \chi^2$ is the $p$-value:
\begin{equation}
    p =
    1 - P_{\chi^2}(\Delta \chi^2, \Delta k = 2) =
    \exp(- \Delta \chi^2/2).
\end{equation}
The $p$-value can then be translated into a number of standard deviations $n_\sigma$ by using the inverse of the standard normal distribution's cumulative distribution function,
\begin{equation}
    n_\sigma =
    \Phi^{-1}(1 - p/2)
    ,\quad
    \Phi(x) =
    \frac{1}{\sqrt{2\pi}} \int_{-\infty}^x e^{-y^2/2}\, {\rm d}y.
\end{equation}

Table~\ref{tab:priors} in Appendix~\ref{app:constraints} lists the priors employed for the sampled parameters, those being a subset of $\boldsymbol{\theta} = (\Omega_{\rm m}, w_0, w_a, \varepsilon)$ depending on the background considered. We also sample the nuisance parameters $\nu$ for SN, $H_0 r_{\rm d}$ when employing BAO in absence of CC, and $H_0$ and $r_{\rm d}$ when combining BAO and CC. Sampling is performed using \textsc{Nautilus}~\cite{Lange:2023ydq} with the standard values of its hyperparameters, except for $\texttt{n}_\texttt{live} = 3,\!000$ and $\texttt{discard\_exploration} = \texttt{true}$. The constraints are reported as $16^{\rm th}$ and $84^{\rm th}$ percentiles about the median, and the $\chi_{\min}^2$ reported throughout and their derived statistics are determined through optimization.

\section{\label{sec:BAO_SN_results}Results and sensitivity of the signal}

We now present the results obtained when combining the uncalibrated BAO+SN samples; unless otherwise specified, tables and figures containing just the SN sample label are understood to be combined with the DESI DR2 data. We first analyze the four models using the datasets as provided to quantify the reduction fraction after freeing the tilt $\varepsilon$. Appendix~\ref{app:constraints} contains the full one-dimensional marginalized constraints for $\boldsymbol{\theta} = (\Omega_{\rm m}, w_0, w_a, \varepsilon)$ in Table~\ref{tab:posterior}, as well as its complementary triangle plot in Fig.~\ref{fig:2x2_posterior}.  Afterwards, we test the robustness of these results under three independent scenarios: (i) alternative $\eta(z)$ parametrizations and the addition of spatial curvature $\Omega_k$; (ii) deletion of different BAO tracers from the DESI DR2 catalog; and (iii) various redshift cutoffs in the Pantheon$+$ and DES-Dovekie samples. All figures presented throughout were produced using \textsc{Matplotlib}~\cite{Hunter:2007}, and the one- and two-dimensional marginalized posteriors with \textsc{GetDist}~\cite{Lewis:2019xzd}.

\subsection{\label{sec:results}Baseline results}

Table~\ref{tab:baseline_improvements} shows that allowing a common tilt reduces the $w_0w_a{\rm CDM}$ preference over $\Lambda{\rm CDM}$ for every SN compilation, with the significance being largely dependent on the supernovae sample. Attending to the reduction fraction $f_{\rm red}$, the preference for $w_0w_a{\rm CDM}$ is reduced by $57\%$ for Pantheon$+$, $43\%$ for both the DES-Dovekie and Union3 combinations, and $18\%$ for the Union3.1. In terms of the number of standard deviations, the preference for dynamical dark energy is reduced from $1.8 \sigma$ to $1.0\sigma$ for Pantheon$+$, $2.2\sigma$ to $1.5\sigma$ for DES-Dovekie, $2.7\sigma$ to $1.9\sigma$ for Union3, and $2.0\sigma$ to $1.8\sigma$ for Union3.1. Values for $\Delta \chi_{{\rm DE}\vert\varepsilon}^2$ across SN samples are comparable to the $\Delta \chi_{\rm DE}^2$ found using BAO only, except for Pantheon$+$, for which it is smaller. In terms of the $\Delta {\rm AIC}_{{\rm DE}\vert\varepsilon}$, the differences are small enough to not provide preference for one background or the other when allowing the common tilt.

\begin{table}[t]
\caption{Improvement of $w_0w_a{\rm CDM}$ relative to $\Lambda{\rm CDM}$ before and after freeing a common tilt $\varepsilon$, quantified by $\Delta \chi_{\rm DE}^2$ and $\Delta \chi_{{\rm DE} \vert \varepsilon}^2$ respectively. Rows quoting the SN are understood to be combined with the DESI DR2 BAO data. We also quote the approximate Gaussian-equivalent significances, calculated from the unrounded likelihood improvements, together with the reduction fraction $f_{\rm red}$ as defined in Eq.~\eqref{eq:reduction_fraction}. For reference, the improvement of $w_0w_a{\rm CDM}$ is also reported for the BAO alone, which are independent of $\varepsilon$.}\label{tab:baseline_improvements}

\begin{ruledtabular}
\begin{tabular}{lccc}
    Sample & $\Delta \chi_{\rm DE}^2 / n_\sigma$ & $\Delta \chi_{{\rm DE} \vert \varepsilon}^2 / n_\sigma$ & $f_{\rm red}(\%)$\vspace{0.1em}\\
\hline\rule{0pt}{1.2em}%
    BAO only & ${\phantom 0}4.65$ / $1.7$ & ------ & ------ \\
    Pantheon$+$ & ${\phantom 0}5.06$ / $1.8$ & $2.18$ / $1.0$ & $57$ \\
    DES-Dovekie & ${\phantom 0}7.16$ / $2.2$ & $4.08$ / $1.5$ & $43$ \\
    Union3 & $10.03$ / $2.7$ & $5.75$ / $1.9$ & $43$ \\
    Union3.1 & ${\phantom 0}6.42$ / $2.0$ & $5.26$ / $1.8$ & $18$ \\
\end{tabular}
\end{ruledtabular}
\end{table}

\begin{figure}[t]
\includegraphics{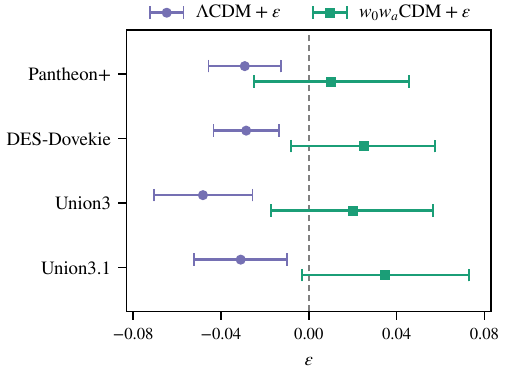}
\caption{Median and $16^{\rm th}$ and $84^{\rm th}$ percentile uncertainty intervals for the tilt $\varepsilon$ for $\Lambda{\rm CDM} + \varepsilon$ and $w_0w_a{\rm CDM} + \varepsilon$. All SN entries are understood to be combined with DESI DR2 BAO. The vertical dashed line corresponds to no CDDR deformation ($\varepsilon = 0$).}\label{fig:epsilon}
\end{figure}

In the $\Lambda{\rm CDM} + \varepsilon$ model, the one-dimensional marginalized posteriors consistently favor a negative tilt $\varepsilon$ for all four supernova compilations. We find $\varepsilon = -0.029 \pm 0.017$ for Pantheon$+$, $-0.029 \pm 0.015$ for DES-Dovekie, $-0.048 \pm 0.023$ for Union3, and $-0.031 \pm 0.021$ for Union3.1, all compatible with no redshift-dependent CDDR tilt $\varepsilon$ within $2.1\sigma$. Once we allow for dynamical dark energy by letting $(w_0, w_a)$ vary freely together with $\varepsilon$, the posterior median becomes positive for all SN compilations. Because of the additional degrees of freedom, the credible intervals increase accordingly; specifically, $\varepsilon = 0.010 \pm 0.035$ for Pantheon$+$, $0.025 \pm 0.033$ for DES-Dovekie, $0.020 \pm 0.037$ for Union3, and $0.035 \pm 0.038$ for Union3.1. A recent analysis including compressed CMB information~\cite{Bansal:2025ipo} instead reported negative central tilts $\varepsilon$ for $w_0w_a{\rm CDM}$~\cite{Dhawan:2025mer}. As the datasets, calibration treatment, and model assumptions also differ, this comparison does not necessarily isolate the effect of adding CMB data. In any case, all these are compatible with $\varepsilon = 0$ within $1\sigma$~\cite{Dhawan:2025mer}. Fig.~\ref{fig:epsilon} summarizes the background dependence of $\varepsilon$.

\begin{figure}[t]
\includegraphics{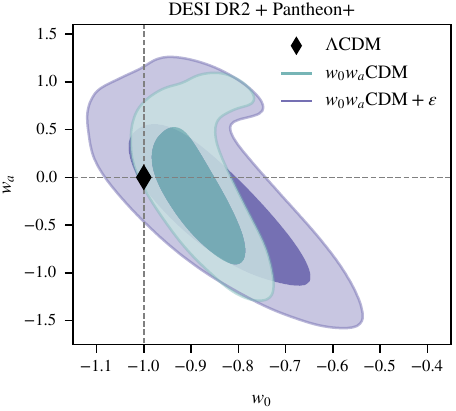}
\caption{Marginalized two-dimensional $(w_0, w_a)$ posteriors for DESI DR2 BAO combined with Pantheon$+$, with $\varepsilon$ set to zero (turquoise) and free $\varepsilon$ (purple). The contours enclose $68\%$ and $95\%$ of the two-dimensional posterior probabilities. The diamond marks the $\Lambda{\rm CDM}$ point, and the dashed lines mark the $w_0 \!=\! -1$ and $w_a \!=\! 0$ values.}\label{fig:w0_wa}
\end{figure}

Regarding the present-day equation of state $w_0$ (see Table~\ref{tab:posterior} for all details), the values remain remarkably stable for Pantheon$+$ before and after letting $\varepsilon$ vary freely, with the one-dimensional confidence intervals increasing approximately by a factor of $2$; for the remaining SN catalogs, we find a more appreciable shift in their values though with a smaller increase in the confidence intervals. The sign flip in $\varepsilon$ when moving from $\Lambda{\rm CDM}$ to $w_0w_a{\rm CDM}$ is a direct consequence of the $w_0$--$\varepsilon$ degeneracy (see Fig.~\ref{fig:2x2_posterior} for details). A similar shift is observed for the $w_a$ parameter, but the errors remain stable when extending the $\varepsilon = 0$ scenario to the free $\varepsilon$ setting. Fig.~\ref{fig:w0_wa} illustrates the two-dimensional $(w_0, w_a)$ marginalized posterior distribution for BAO$+$Pantheon$+$ before and after freeing $\varepsilon$. The addition of $\varepsilon$ slightly increases the anticorrelation between $w_0$ and $w_a$, but greatly increases the confidence regions. This is mainly caused by the increased uncertainty in $w_0$, which reduces the distance to the $\Lambda$ point in this two-dimensional marginalized parameter space from $\sim\!2\sigma$ to $\sim\!1\sigma$.

We also compare the non-nested models $\Lambda{\rm CDM} + \varepsilon$ and $w_0w_a{\rm CDM}$ using $\Delta \chi^2$ and $\Delta {\rm AIC}$, with positive values for these favoring $w_0w_a{\rm CDM}$. Table~\ref{tab:non-nested} reports these statistics for the four BAO$+$SN combinations. $w_0w_a{\rm CDM}$ gives smaller $\chi_{\rm min}^2$ in every case, though the additional parameter with respect to $\Lambda{\rm CDM} + \varepsilon$ reduces the AIC improvement for Pantheon$+$, where $\Delta{\rm AIC} \simeq 0$. The other AIC differences are $1.36$ for DES-Dovekie, $3.40$ for Union3, and $2.25$ for Union3.1, corresponding to probabilities of $66\%$, $85\%$, and $75\%$, respectively, in favor of CPL over $\Lambda{\rm CDM} + \varepsilon$.

\subsection{Dependence on parametrizations and data selection}

\subsubsection{Alternative magnitude tilts and geometries}

The distance-duality parametrization~\eqref{eq:eta} produces an apparent magnitude or distance modulus evolution with redshift for SN of $\Delta m = \Delta \mu = 5\varepsilon \log_{10}(1+z)$. However, one can consider alternative perturbations with redshift which could alter the significances previously reported (see e.g.~Ref.~\cite{Sapone:2020wwz} for earlier implementations). To address this question, we consider a linear perturbation $\Delta m = \kappa z$ and a bounded, CPL-like perturbation $\Delta m = \lambda z/(1+z)$.%
\footnote{The corresponding effective CDDR parameters would be $\eta_\kappa(z) = \alpha 10^{\kappa z/5}$ and $\eta_\lambda(z) = \alpha 10^{\lambda z/[5(1+z)]}$, respectively.}
As before, the same perturbation is allowed in both the $\Lambda{\rm CDM}$ and the $w_0w_a{\rm CDM}$ backgrounds. Separately, we also allow spatial curvature $\Omega_k$ to vary freely (fixing the tilt to zero) and quantify the reduction in preference it produces. Note that the parameters entering $\eta(z)$ affect only the SN, whereas $\Omega_k$ influences both BAO and SN distances.

\begin{table}[t]
\caption{Model comparison between the non-nested models $\Lambda{\rm CDM} + \varepsilon$ and $w_0w_a{\rm CDM}$. Rows quoting the SN are understood to be combined with the DESI DR2 BAO data. Here $\Delta \chi^2 \equiv \min(\chi_{\Lambda{\rm CDM} + \varepsilon}^2) - \min(\chi_{w_0w_a{\rm CDM}}^2)$, with $\Delta {\rm AIC}$ following the same criteria: positive values favor $w_0w_a{\rm CDM}$ over $\Lambda{\rm CDM} + \varepsilon$.}\label{tab:non-nested}
\begin{ruledtabular}
\begin{tabular}{clccc}
& Sample & $\Delta \chi^2$ & $\Delta{\rm AIC}$ &\\
\hline
& Pantheon$+$ &  $2.00$ & $0.00$ & \\
& DES-Dovekie &  $3.36$ & $1.36$ & \\
& Union3 &  $5.40$ & $3.40$ & \\
& Union3.1 &  $4.25$ & $2.25$ & \\
\end{tabular}
\end{ruledtabular}
\end{table}

\begin{figure}[t]
    \includegraphics{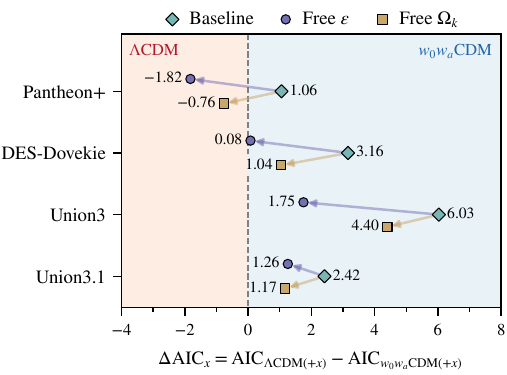}
    \caption{${\rm AIC}$ differences between $\Lambda{\rm CDM}$ and $w_0w_a{\rm CDM}$ for DESI DR2 BAO plus each SN sample in three settings: no common extension (baseline), a common free tilt $\varepsilon$ (free $\varepsilon$), and common free spatial curvature $\Omega_k$ with no tilt (free $\Omega_k$). Positive values in the blue region favor $w_0w_a{\rm CDM}$, whereas negative values in the red region favor $\Lambda{\rm CDM}$. A difference $\Delta {\rm AIC}_x$ corresponds to a probability $[1 + \exp(-\vert \Delta {\rm AIC}_x \vert/2)]^{-1}$ for $w_0w_a{\rm CDM}$ in terms of Akaike weights. Arrows connect the baseline to each one-parameter extension; vertical offsets are added for readability only.}\label{fig:comparison}
\end{figure}

Table~\ref{tab:controls} contains the relative improvement in the fits of $w_0w_a{\rm CDM}$ with respect to $\Lambda{\rm CDM}$ for each of the one-parameter extensions mentioned. In this case, $\Delta \chi_{{\rm DE}\vert x}^2$ with $x$ denoting $\{\varepsilon, \kappa, \lambda, \Omega_k\}$ are the likelihood differences as defined in Eq.~\eqref{eq:delta_chi2_de_eps} but for the corresponding parameter extensions $x$. Note also that larger values of $\Delta \chi_{{\rm DE}\vert x}^2$ correspond to smaller reduction fractions $f_{\rm red}$. As shown in Table~\ref{tab:controls}, $\Omega_k$ produces the smallest reduction across all one-parameter extensions for all datasets, whereas $\varepsilon$, $\kappa$, or $\lambda$ produce comparable reductions within the same data combination. Union3.1 is the only exception, for which any of the one-parameter extensions considered performs equally well in absorbing a small portion of the dynamical dark energy signal. Fig.~\ref{fig:comparison} compares the improvement with respect to the baseline in terms of $\Delta{\rm AIC}$ when freeing $\varepsilon$ or $\Omega_k$ in both $\Lambda{\rm CDM}$ and $w_0w_a{\rm CDM}$. With a common free tilt, the $\Delta {\rm AIC}_{{\rm DE}\vert\varepsilon}$ are $-1.82$, $0.08$, $1.75$, and $1.26$ for Pantheon$+$, DES-Dovekie, Union3, and Union3.1, whereas the corresponding curvature-extension differences are $-0.76$, $1.04$, $4.40$, and $1.17$: the $\varepsilon$ extension significantly reduces the preference for dynamical dark energy, and it does so in a more efficient manner than other non CDDR violating parameters like $\Omega_k$. Overall, $\varepsilon$ constitutes a useful low-dimensional test, encapsulating the relevant information from CDDR deviations.

\begin{table}[t]
\caption{Likelihood improvement of $w_0w_a{\rm CDM}$ with respect to $\Lambda{\rm CDM}$ for different common extensions. Rows quoting the SN are understood to be combined with the DESI DR2 BAO data. The logarithmic, linear, and CPL-like columns ($\varepsilon$, $\kappa$, and $\lambda$, respectively) allow the indicated SN magnitude perturbation in both models. The last column frees $\Omega_k$ in both backgrounds and sets the tilt to zero.}\label{tab:controls}

\begin{ruledtabular}
\begin{tabular}{lccccc}
Sample & $\Delta \chi_{\rm DE}^2$ & $\Delta \chi_{{\rm DE}\vert\varepsilon}^2$ & $\Delta \chi_{{\rm DE}\vert\kappa}^2$ & $\Delta \chi_{{\rm DE}\vert\lambda}^2$ & $\Delta \chi_{{\rm DE}\vert\Omega_k}^2$\vspace{0.1em}\\
\hline\rule{0pt}{1.25em}%
Pantheon$+$ & ${\phantom 0}5.06$ & $2.18$ & $2.64$ & $2.05$ & $3.24$ \\
DES-Dovekie & ${\phantom 0}7.16$ & $4.08$ & $4.16$ & $4.06$ & $5.04$ \\
Union3 & $10.03$ & $5.75$ & $6.47$ & $5.20$ & $8.40$ \\
Union3.1 & ${\phantom 0}6.42$ & $5.26$ & $5.29$ & $5.26$ & $5.17$ \\
\end{tabular}
\end{ruledtabular}
\end{table}

\begin{table*}[t]
\caption{DESI DR2 single-tracer deletions for all four supernova likelihoods. For each dropped tracer we quote the improvement in fit of $w_0w_a{\rm CDM}$ over $\Lambda{\rm CDM}$ with $\varepsilon=0$, $\Delta\chi^2_{\rm DE}$, and with $\varepsilon$ left free in both models, $\Delta\chi^2_{{\rm DE}\vert\varepsilon}$, together with the reduction fraction $f_{\rm red}$ from Eq.~\eqref{eq:reduction_fraction}. Larger values of $f_{\rm red}$ imply a reduced preference for $w_0w_a{\rm CDM}$ compared with $\Lambda{\rm CDM}$ once $\varepsilon$ is left to vary freely in both models.}\label{tab:bao_jackknife_all}
\begin{ruledtabular}
\begin{tabular}{lccccccccccccccc}
 & \multicolumn{3}{c}{Pantheon$+$} & & \multicolumn{3}{c}{DES-Dovekie} & & \multicolumn{3}{c}{Union3} & & \multicolumn{3}{c}{Union3.1}\\
\cline{2-4}\cline{6-8}\cline{10-12}\cline{14-16}\rule{0pt}{1.2em}%
Dropped tracer
 & $\Delta\chi^2_{\rm DE}$ & $\Delta\chi^2_{{\rm DE}\vert\varepsilon}$ & $f_{\rm red}(\%)$ &
 & $\Delta\chi^2_{\rm DE}$ & $\Delta\chi^2_{{\rm DE}\vert\varepsilon}$ & $f_{\rm red}(\%)$ &
 & $\Delta\chi^2_{\rm DE}$ & $\Delta\chi^2_{{\rm DE}\vert\varepsilon}$ & $f_{\rm red}(\%)$ &
 & $\Delta\chi^2_{\rm DE}$ & $\Delta\chi^2_{{\rm DE}\vert\varepsilon}$ & $f_{\rm red}(\%)$\vspace{0.1em}\\
\hline\rule{0pt}{1.2em}%
None (full DR2)      & $5.06$ & $2.18$ & $57$ && $7.16$ & $4.08$ & $43$ && $10.03$ & $5.75$ & $43$ && $6.42$ & $5.26$ & $18$ \\
\texttt{BGS}         & $5.06$ & $1.82$ & $64$ && $7.34$ & $3.69$ & $50$ && ${\phantom 0}9.99$  & $5.41$ & $46$ && $6.41$ & $4.89$ & $24$ \\
\texttt{LRG1}        & $5.14$ & $1.25$ & $76$ && $7.86$ & $3.17$ & $60$ && ${\phantom 0}9.97$  & $4.64$ & $54$ && $6.54$ & $4.20$ & $36$ \\
\texttt{LRG2}        & $3.69$ & $1.07$ & $71$ && $5.89$ & $2.87$ & $51$ && ${\phantom 0}8.20$  & $4.04$ & $51$ && $5.12$ & $3.67$ & $28$ \\
\texttt{LRG3$+$ELG1} & $5.29$ & $3.23$ & $39$ && $6.60$ & $4.72$ & $28$ && ${\phantom 0}9.86$  & $6.33$ & $36$ && $6.21$ & $5.82$ & ${\phantom 0}6$  \\
\texttt{ELG2}        & $4.71$ & $2.17$ & $54$ && $6.57$ & $3.94$ & $40$ && ${\phantom 0}9.52$  & $5.53$ & $42$ && $6.02$ & $5.05$ & $16$ \\
\texttt{QSO}         & $5.00$ & $2.20$ & $56$ && $7.01$ & $4.04$ & $42$ && ${\phantom 0}9.89$  & $5.68$ & $43$ && $6.30$ & $5.19$ & $18$ \\
\texttt{Ly$\alpha$}  & $5.68$ & $3.71$ & $35$ && $6.22$ & $4.47$ & $28$ && ${\phantom 0}9.53$  & $6.00$ & $37$ && $6.19$ & $5.48$ & $11$ \\
\end{tabular}
\end{ruledtabular}
\end{table*}

\subsubsection{\label{sec:bao_tracer_deletion}BAO tracer deletion}

We now test the sensitivity to individual BAO tracers by removing each of the (uncorrelated) seven released DESI DR2 blocks in turn. BGS removes one $D_{\rm V}/r_{\rm d}$ data point, whereas every other tracer removes its correlated $D_{\rm M}/r_{\rm d}$ and $D_{\rm H}/r_{\rm d}$ pair. For each deletion, we select the principal submatrix of the covariance matrix and invert it, and refit all four models.

Table~\ref{tab:bao_jackknife_all} gives $\Delta\chi^2_{\rm DE}$, $\Delta\chi^2_{{\rm DE}\vert\varepsilon}$, and $f_{\rm red}$ for each deletion. We confirm that the best-fit tilts $\varepsilon$ in the extended $\Lambda{\rm CDM} + \varepsilon$ and $w_0w_a{\rm CDM} + \varepsilon$ remained stable, although we omit reporting the values for brevity. The reduction fraction increases when the low-redshift BGS, LRG1, and LRG2 blocks are removed, decreases when LRG3+ELG1 or Ly$\alpha$ are removed, and remains nearly unchanged for ELG2 and QSO, with the same direction of change for all four SN samples. In most cases, these shifts are driven by $\Delta\chi^2_{{\rm DE}\vert\varepsilon}$ rather than by the baseline preference $\Delta \chi_{\rm DE}^2$. LRG2 is the only block whose removal lowers both statistics by comparable amounts. The deletions therefore mainly probe how much of the $w_0w_a{\rm CDM}$ preference survives once $\varepsilon$ is free, given the retained BAO redshifts: LRG1 and LRG2 support this residual preference, while LRG3+ELG1 and Ly$\alpha$ counteract it.%
\footnote{For the Pantheon$+$Ly$\alpha$ deletion, the $w_0w_a{\rm CDM}$ likelihoods are nearly flat along the $\Omega_{\rm m}$ density, and both minimizations reach the lower prior boundary $\Omega_{\rm m} = 0.1$ (see Table~\ref{tab:priors}). Nonetheless, lowering this bound to $0.001$ changes $f_{\rm red}$ by less than $0.5\%$, and therefore the likelihood is stable under this extension. Given the weak determination of $\Omega_{\rm m}$, we opt for not assigning Wilks significances to any of the values in Table~\ref{tab:bao_jackknife_all}.}
We stress that these fits share almost all their data, so the reduction fractions are descriptive rather than independent significances, and we do not conclude that any particular block acts as an outlier or that the fitted tilt $\varepsilon$ has a physical origin.

\subsubsection{\label{sec:low_redshift_cutoff}Low-redshift cutoffs on SN}

We next quantify the impact of removing low-redshift data points in the uncompressed Pantheon$+$ and DES-Dovekie samples. We do this by removing their data points and slicing the covariance matrix to the respective block. We leave both Union3 and Union3.1 samples untouched, as these constitute a compression of a SN catalog for which no clear redshift cutoff can be imposed (unless a reanalysis of the full data is performed, which lies outside the scope of this work).

\begin{table}[t]
\caption{Reanalysis of BAO plus either Pantheon$+$ or DES-Dovekie after removing the samples with $z_{\rm HD} < z_{\min}$. For Pantheon$+$, the cutoff $z_{\rm min} = 0.023$ is selected as a representative SH0ES Hubble-flow threshold~\cite{Riess:2021jrx}. The DES-Dovekie catalog already supplies a selection of SN of $z > 0.025$, and so the first row corresponds to the baseline results presented in Sec.~\ref{sec:results}. The tilts $\varepsilon_{\Lambda{\rm CDM}}$ correspond to the median and 16/84th percentiles for the $\Lambda{\rm CDM} + \varepsilon$ model, $N_{\rm SN}$ denotes the remaining number of SN after the redshift cutoff, and $\Delta \chi_{\rm DE}^2$ and $\Delta \chi_{{\rm DE}\vert \varepsilon}^2$ are the likelihood improvements defined in Eqs.~\eqref{eq:delta_chi2_de} and~\eqref{eq:delta_chi2_de_eps}, respectively.}\label{tab:cuts}

\begin{ruledtabular}
\begin{tabular}{lcrccc}
Sample & $z_{\min}$ & $N_{\rm SN}$ & $\varepsilon_{\Lambda{\rm CDM}}$ & $\Delta \chi_{\rm DE}^2$ & $\Delta \chi_{{\rm DE}\vert \varepsilon}^2$ \vspace{0.1em} \\ \hline\rule{0pt}{1.2em}%
Pantheon$+$ & $0.010$ & 1580 & $-0.029 \pm 0.017$ & $5.06$ & $2.18$ \\
 & $0.023$ & 1371 & $-0.025 \pm 0.017$ & $3.58$ & $1.44$ \\
 & $0.050$ & 1056 & $-0.029 \pm 0.019$ & $3.40$ & $1.19$ \\
 & $0.100$ & 960 & $-0.029 \pm 0.022$ & $3.21$ & $1.44$ \\
\hline\rule{0pt}{1.2em}%
DES-Dovekie & $0.025$ & 1820 & $-0.029 \pm 0.015$ & $7.16$ & $4.08$ \\
 & $0.050$ & 1679 & $-0.021 \pm 0.017$ & $3.24$ & $1.89$ \\
 & $0.100$ & 1619 & $-0.008 \pm 0.020$ & $0.95$ & $0.82$ \\
\end{tabular}
\end{ruledtabular}
\end{table}

Table~\ref{tab:cuts} contains the $\varepsilon$ tilts for the $\Lambda{\rm CDM} + \varepsilon$ model and both dynamical dark energy improvements $\Delta \chi_{\rm DE}^2$ and $\Delta \chi_{{\rm DE} \vert \varepsilon}^2$ before and after letting $\varepsilon$ vary freely. The Pantheon$+$ tilts $\varepsilon$ remain near $-0.03$, with increasing uncertainty as the number of removed SN increases. For DES-Dovekie, removing the low-redshift part (which substantially overlaps with those in the Pantheon$+$ sample~\cite{DES:2025sig}) weakens both the negative tilt and the $w_0w_a{\rm CDM}$ improvement. Specifically, for $z_{\rm min} = 0.1$ the tilt is consistent with no redshift-evolution of CDDR within $1\sigma$, and the dark-energy improvement relative to $\Lambda{\rm CDM}$ drops significantly for both $\varepsilon = 0$ and free $\varepsilon$ scenarios.%
\footnote{For DES-Dovekie at $z_{\rm min} = 0.1$, both $w_0w_a{\rm CDM}$ and $w_0w_a{\rm CDM} + \varepsilon$ lie at the adopted lower prior boundary $\Omega_{\rm m} = 0.1$. Table~\ref{tab:cuts} uses these lower minima, whereas solutions in the interior of the priors yield $\Delta \chi_{\rm DE}^2 = 0.63$ and $\Delta \chi_{{\rm DE}\vert\varepsilon}^2 = 0.44$; the conclusion that the CPL preference remains small after this cutoff remains. We opt for not assigning Wilks significances to the values reported in Table~\ref{tab:cuts}.}
This behavior matches the low-$z$ offset reported for DES-SN5YR~\cite{Efstathiou:2024xcq} (even though DES-Dovekie constitutes a reanalysis). This is evidence that the available redshift range and its relative calibration matter. In any case, this does not demonstrate that a particular low-redshift systematic caused this result, and these cutoffs do not erase the influence of low-redshift information already used in the light-curve training and standardization of the samples.

\begin{table*}[t]
\caption{One-dimensional marginalized posterior medians and 16th/84th-percentile errors for DESI DR2 BAO plus each SN compilation and CC.\@ The first-column labels only denote the SN sample, but are all understood to be combined with BAO and CC. Here $H_0$ is in units of ${\rm km}\, {\rm s}^{-1}\, {\rm Mpc}^{-1}$ and $r_{\rm d}$ in units of Mpc. The $\chi_{\min}^2$ column contains separately optimized likelihood minima.}\label{tab:posteriorCC}
\footnotesize
\begin{ruledtabular}
\begin{tabular}{llrcccccc}
Sample & Model & $\chi^2_{\min}$ & $\Omega_{\rm m}$ & $w_0$ & $w_a$ & $\varepsilon$ & $H_0$ & $r_{\rm d}$ \\
\hline\rule{0pt}{1.4em}%
Pantheon$+$ & $\Lambda{\rm CDM}$ & $1414.51$ & $0.3050^{+0.0079}_{-0.0078}$ & $-1$ & $0$ & $0$ & $68.8^{+3.3}_{-3.2}$ & $146.7^{+7.1}_{-6.7}$ \\[0.37em]
 & $w_0w_a{\rm CDM}$ & $1409.63$ & $0.303^{+0.016}_{-0.029}$ & $-0.888^{+0.064}_{-0.057}$ & $-0.14^{+0.55}_{-0.46}$ & $0$ & $67.9^{+3.2}_{-3.3}$ & $146.7^{+7.4}_{-6.6}$ \\[0.37em]
 & $\Lambda{\rm CDM} + \varepsilon$ & $1411.54$ & $0.2983^{+0.0085}_{-0.0084}$ & $-1$ & $0$ & $-0.029^{+0.016}_{-0.016}$ & $69.2^{+3.3}_{-3.3}$ & $146.6^{+7.3}_{-6.7}$ \\[0.37em]
 & $w_0w_a{\rm CDM} + \varepsilon$ & $1409.50$ & $0.305^{+0.021}_{-0.033}$ & $-0.86^{+0.11}_{-0.11}$ & $-0.22^{+0.66}_{-0.54}$ & $0.008^{+0.035}_{-0.034}$ & $67.6^{+3.4}_{-3.4}$ & $146.9^{+7.4}_{-6.5}$ \\[0.37em]
\hline\rule{0pt}{1.4em}%
DES-Dovekie & $\Lambda{\rm CDM}$ & $1659.98$ & $0.3065^{+0.0075}_{-0.0075}$ & $-1$ & $0$ & $0$ & $68.7^{+3.4}_{-3.2}$ & $146.9^{+7.2}_{-6.9}$ \\[0.37em]
 & $w_0w_a{\rm CDM}$ & $1653.12$ & $0.312^{+0.014}_{-0.019}$ & $-0.850^{+0.070}_{-0.067}$ & $-0.47^{+0.51}_{-0.47}$ & $0$ & $67.9^{+3.3}_{-3.2}$ & $146.6^{+7.1}_{-6.6}$ \\[0.37em]
 & $\Lambda{\rm CDM} + \varepsilon$ & $1656.28$ & $0.2978^{+0.0084}_{-0.0083}$ & $-1$ & $0$ & $-0.028^{+0.015}_{-0.015}$ & $69.2^{+3.4}_{-3.4}$ & $146.6^{+7.5}_{-6.6}$ \\[0.37em]
 & $w_0w_a{\rm CDM} + \varepsilon$ & $1652.49$ & $0.322^{+0.018}_{-0.022}$ & $-0.78^{+0.13}_{-0.12}$ & $-0.67^{+0.55}_{-0.55}$ & $0.023^{+0.032}_{-0.032}$ & $66.9^{+3.4}_{-3.4}$ & $146.6^{+7.4}_{-6.6}$ \\[0.37em]
\hline\rule{0pt}{1.4em}%
Union3 & $\Lambda{\rm CDM}$ & $53.54$ & $0.3044^{+0.0083}_{-0.0082}$ & $-1$ & $0$ & $0$ & $68.9^{+3.3}_{-3.4}$ & $146.5^{+7.6}_{-6.6}$ \\[0.37em]
 & $w_0w_a{\rm CDM}$ & $43.99$ & $0.330^{+0.016}_{-0.018}$ & $-0.71^{+0.11}_{-0.11}$ & $-0.95^{+0.57}_{-0.56}$ & $0$ & $66.3^{+3.4}_{-3.2}$ & $146.9^{+7.3}_{-6.8}$ \\[0.37em]
 & $\Lambda{\rm CDM} + \varepsilon$ & $48.98$ & $0.2981^{+0.0087}_{-0.0084}$ & $-1$ & $0$ & $-0.047^{+0.022}_{-0.023}$ & $69.2^{+3.3}_{-3.4}$ & $146.5^{+7.5}_{-6.6}$ \\[0.37em]
 & $w_0w_a{\rm CDM} + \varepsilon$ & $43.73$ & $0.336^{+0.022}_{-0.024}$ & $-0.66^{+0.17}_{-0.17}$ & $-1.11^{+0.69}_{-0.67}$ & $0.016^{+0.037}_{-0.037}$ & $65.8^{+3.5}_{-3.4}$ & $147.0^{+7.2}_{-6.8}$ \\[0.37em]
\hline\rule{0pt}{1.4em}%
Union3.1 & $\Lambda{\rm CDM}$ & $55.83$ & $0.3024^{+0.0082}_{-0.0083}$ & $-1$ & $0$ & $0$ & $68.9^{+3.3}_{-3.3}$ & $146.7^{+7.4}_{-6.6}$ \\[0.37em]
 & $w_0w_a{\rm CDM}$ & $49.78$ & $0.320^{+0.016}_{-0.020}$ & $-0.79^{+0.11}_{-0.10}$ & $-0.70^{+0.58}_{-0.56}$ & $0$ & $67.1^{+3.3}_{-3.3}$ & $146.8^{+7.4}_{-6.6}$ \\[0.37em]
 & $\Lambda{\rm CDM} + \varepsilon$ & $53.72$ & $0.2977^{+0.0088}_{-0.0081}$ & $-1$ & $0$ & $-0.031^{+0.021}_{-0.021}$ & $69.2^{+3.4}_{-3.3}$ & $146.7^{+7.4}_{-6.7}$ \\[0.37em]
 & $w_0w_a{\rm CDM} + \varepsilon$ & $48.92$ & $0.334^{+0.021}_{-0.024}$ & $-0.68^{+0.16}_{-0.16}$ & $-1.04^{+0.67}_{-0.63}$ & $0.032^{+0.037}_{-0.037}$ & $65.9^{+3.6}_{-3.5}$ & $147.0^{+7.4}_{-6.7}$ \\[0.37em]
\end{tabular}
\end{ruledtabular}
\end{table*}

\section{\label{sec:Hubble_tension}Distance duality and Hubble tension}

In this section we consider two complementary extensions to the uncalibrated BAO+SN analysis. We first start by adding cosmic chronometers (CC) to the previous analysis as described in Sec.~\ref{sec:datasets} in order to break the $H_0$--$r_{\rm d}$ degeneracy. This allows us to assess whether the four models and four supernova samples considered reveal a tension between the Hubble constant $H_0$ inferred from low-redshift Cepheid anchors and that inferred from the CMB.\@ Second, we employ illustrative Gaussian priors on the sound horizon $r_{\rm d}$ from Planck~\cite{Planck:2018vyg} and on the absolute magnitude $M_{\rm B}$ from SH0ES~\cite{Riess:2021jrx} to provide calibrated constraints on the normalization parameter $\alpha$ in Eq.~\eqref{eq:eta}; the amount by which $\alpha$ deviates from unity quantifies the requirement to reconcile early- and late-time constraints on $H_0$. For a non-exhaustive selection of discussion on this topic, we refer the reader to Refs.~\cite{Poulin:2024ken,Teixeira:2025czm, Kanodia:2025jqh, Afroz:2025iwo, Tiwari:2026pzk, Bansal:2026axl, Martinelli:2026wjp}.

\subsection{\label{sec:cosmic_chronometer_calibration}Cosmic chronometer calibration}

In the adopted effective model, both BAO and CC are insensitive to the tilt $\varepsilon$, which acts solely on the SN luminosity distances. Nonetheless, their joint combination allows us to break the $H_0$--$r_{\rm d}$ degeneracy in the absence of a sound-horizon prior coming from other data, e.g.~CMB.\@ For the $\Lambda{\rm CDM}$ model, BAO$+$CC give $H_0 = 69.2 \pm 3.3\, {\rm km}\, {\rm s}^{-1}\, {\rm Mpc}^{-1}$ and $r_{\rm d} = 146.6_{-6.6}^{+7.4}\, {\rm Mpc}$. Both constraints are compatible with the Planck-calibrated values, but also with the SN calibrated values $r_{\rm d} \simeq 139\, {\rm Mpc}$ and $H_0 \simeq 73\, {\rm km}\, {\rm s}^{-1}\, {\rm Mpc}^{-1}$ within $1.2\sigma$.

The weak discriminatory power of CC is also reflected in the model comparison. For CC alone, the improvement of the $w_0w_a{\rm CDM}$ model is $\Delta \chi_{\rm DE}^2 = 0.22$; BAO alone and BAO$+$CC however increase this preference relative to CC alone to $\Delta \chi_{\rm DE}^2 = 4.65$ and $3.43$, respectively. We also report in Table~\ref{tab:posteriorCC} the full BAO$+$SN$+$CC results. In number of standard deviations, the preference for dynamical dark energy compared with $\Lambda{\rm CDM}$ before and after freeing $\varepsilon$ reduces from $(1.7,\, 2.1,\, 2.6,\, 2.0)\sigma$ to $(0.9,\, 1.4,\, 1.8,\, 1.7)\sigma$ for Pantheon$+$, DES-Dovekie, Union3, and Union3.1, respectively. The parameter constraints, as well as both $\Delta \chi_{\rm DE}^2$ and $\Delta \chi_{{\rm DE}\vert \varepsilon}^2$ present modest differences with the BAO$+$SN baseline analysis in Table~\ref{tab:posterior}. The constraints on $H_0$ and $r_{\rm d}$ remain stable across models, with values compatible with Planck constraints and deviations of at most $\sim2\sigma$ from the sound horizon and Hubble parameter predicted by late-time calibrations. Therefore, current CC measurements cannot establish whether the calibration discrepancy resides in the SN absolute magnitude $M_{\rm B}$ or in the sound horizon.

\begin{figure*}[t]
    \includegraphics{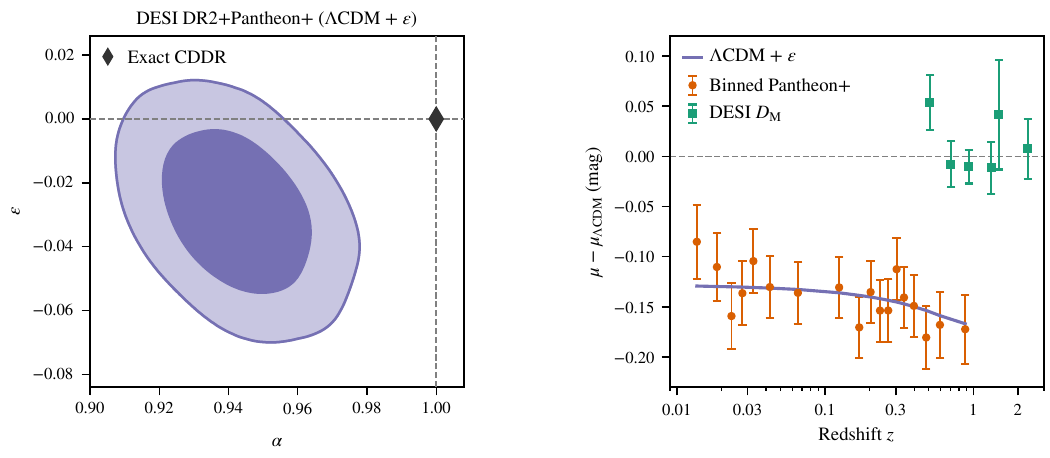}
    \caption{{\bfseries Left:} Marginalized $(\alpha, \varepsilon)$ posterior for the CDDR violating parameters when combining DESI DR2 BAO$+$Pantheon$+$ together with a Planck prior $r_{\rm d} = 147.09 \pm 0.26\, {\rm Mpc}$ and a SH0ES prior $M_{\rm B} = -19.253 \pm 0.027\, {\rm mag}$. Contours enclose $68$ and $95\%$ probability regions; the black diamond marks the exact CDDR point $(\alpha, \varepsilon) = (1,0)$. {\bfseries Right:} Distance modulus residuals relative to a $\Lambda{\rm CDM}$ fit to the full DESI DR2 likelihood. Transverse BAO distances use the Planck sound-horizon prior and standard CDDR; Pantheon$+$ uses SH0ES absolute magnitude prior and $18$ approximately equal-population bins. The bins are correlated and are used only for illustration. The purple curve shows the best-fitting uncalibrated BAO+SN distance modulus with its normalization set by the SH0ES prior. The offset between BAO and SN diagrams illustrates a calibration mismatch, and the bending of $\Lambda{\rm CDM} + \varepsilon$ reflects the fact that $\varepsilon \neq 0$.}\label{fig:calibration}
\end{figure*}

\subsection{\label{sec:calibrated_constraints_on_CDDR}Constraints on the CDDR normalization}

We now combine BAO$+$Pantheon$+$ with independent Gaussian calibrations $r_{\rm d} = 147.09 \pm 0.26\, {\rm Mpc}$ from Planck~\cite{Planck:2018vyg} and $M_{\rm B} = -19.253 \pm 0.027\, {\rm mag}$ from SH0ES~\cite{Riess:2021jrx}. Together, they provide the necessary information needed to constrain $\alpha$. In the free-$\alpha$ inference, we retain the baseline uniform priors on $H_0 r_{\rm d}$ and $\nu$ (see Table~\ref{tab:priors}), include the stated Gaussian calibration priors, and derive $H_0$ and $\alpha$. The results presented here are conditional both on the calibrations and the background adopted (see, however, Refs.~\cite{Koksbang:2026jxb, Koksbang:2026wvh} for an alternative background-free analysis), and should not be interpreted as a calibration-free detection of CDDR violation.

Fitting $\Lambda{\rm CDM}$ (setting $\varepsilon = 0$) separately to BAO and Pantheon$+$ yields $H_0^{\rm BAO} = 69.02 \pm 0.52\, {\rm km}\, {\rm s^{-1}}\, {\rm Mpc}^{-1}$ and $H_0^{\rm SN} = 73.23 \pm 0.95\, {\rm km}\, {\rm s^{-1}}\, {\rm Mpc}^{-1}$, translating into a $3.9\sigma$ discrepancy. This difference can be equivalently mapped onto a magnitude offset $\Delta m \simeq -0.13\, {\rm mag}$~\cite{Efstathiou:2021ocp, Camarena:2023rsd}, which in turn maps onto the CDDR normalization as $\alpha \simeq H_0^{\rm BAO}/H_0^{\rm SN} \simeq 0.94$. This ratio is illustrative because the separate BAO and SN fits need not prefer the same $\Omega_{\rm m}$. To account for the different redshift dependence of BAO and SN distances, we perform a joint inference for the $\Lambda{\rm CDM} + \varepsilon$ model using the $r_{\rm d}$ and $M_{\rm B}$ calibrations from early- and late-time data, respectively. Doing so provides precise one-dimensional marginalized constraints on the parameters of $\eta(z)$:
\begin{equation}
    \alpha = 0.942 \pm 0.014
    ,\quad
    \varepsilon = -0.029 \pm 0.017.
\end{equation}
The offset $\alpha$ deviates from unity by $4.1\sigma$ for the $\Lambda{\rm CDM}$ background and the adopted priors. The joint calibrated two-dimensional marginalized posterior for $(\alpha, \varepsilon)$, as shown in the left panel of Fig.~\ref{fig:calibration}, exhibits a moderate anticorrelation.

The right panel of Fig.~\ref{fig:calibration} visualizes the same scale mismatch relative to a $\Lambda{\rm CDM}$ fit to the full DESI DR2 likelihood. The BAO points are obtained by converting the transverse comoving distance $D_{\rm M}$ to distance moduli using the Planck sound-horizon prior $r_{\rm d} = 147.09 \pm 0.26\, {\rm Mpc}$. On the other hand, the Pantheon$+$ data constitutes a binning of the sample considered throughout onto $18$ data points containing approximately the same number of SN, with magnitudes converted using the SH0ES absolute magnitude prior $M_{\rm B} = -19.253 \pm 0.027$. The bins are used only for visualization purposes. The purple curve displays the best-fit of the $\Lambda{\rm CDM} + \varepsilon$ model with respect to the full uncalibrated and uncompressed BAO$+$Pantheon$+$ data, with its normalization set by the SH0ES prior. Its offset and redshift dependence illustrate the separate roles of the normalization $\alpha$ and the shape $\varepsilon$.

The moderate anticorrelation in the left panel of Fig.~\ref{fig:calibration} suggests that fixing $\alpha = 1$ pushes the tilt further away from zero. Imposing $\alpha = 1$ with the same Gaussian calibration priors on $r_{\rm d}$ and $M_{\rm B}$, BAO$+$Pantheon$+$ in the flat $\Lambda{\rm CDM} + \varepsilon$ model yields $\varepsilon = -0.053 \pm 0.016$, corresponding to an approximate $3.3\sigma$ displacement from $\varepsilon = 0$. This is conditional on the imposed calibrations, and should not be interpreted as strong calibration-independent evidence for a redshift-dependent CDDR violation. With $\eta(z=0) = 1$, the tilt can only partially mimic a constant offset, transferring the remaining mismatch to the SN calibration. Indeed, the posterior shifts to $M_{\rm B} = -19.343 \pm 0.014$, a displacement of $0.09\, {\rm mag}$, or $3.3\sigma$ of the SH0ES prior below its mean. Note that this displacement is not an independent tension measurement, as the posterior already incorporates the $M_{\rm B}$ prior. Accordingly, the inferred $H_0$ increases from $69.01 \pm 0.51$ to $70.05 \pm 0.44\,{\rm km\,s^{-1}\,Mpc^{-1}}$. Ref.~\cite{Dhawan:2025mer} finds a qualitatively related shift toward negative $\varepsilon$ after introducing local SN calibration in a $w_0w_a{\rm CDM}$ analysis that does not account for a free $\alpha$. That comparison involves different datasets and assumptions, and should not be understood as a direct free-$\alpha$ vs.~fixed-$\alpha$ test.

\section{\label{sec:physics}Physical interpretation}

An effective departure in the SN and BAO inferred distances does not identify a unique physical mechanism~\cite{Tiwari:2026pzk}. Here we discuss two illustrative mappings of the tilt $\varepsilon$ explored so far: modified photon propagation and evolution of the standardized SN luminosity through an effective gravitational coupling $G(z)$. We note that neither example below provides an independent explanation for a free constant calibration offset.

\subsection{Photon loss interpretation}

If photon number is not conserved during the propagation from the source to Earth, one can effectively parametrize this effect in terms of a photon-loss optical depth $\tau(z)$ that rescales the flux of the source as $e^{-\tau(z)}$~\cite{Avgoustidis:2010ju}. From the standard definition of luminosity distance in terms of the flux, this optical depth can be related to the CDDR violating parametrization~\eqref{eq:eta} as $\eta(z) = e^{\tau(z)/2}$ which, for our specific realization and after fixing the normalization at $z = 0$, yields
\begin{equation}
    \tau(z) = 2\varepsilon \ln(1+z).
\end{equation}
The negative central tilts $\varepsilon$ for the $\Lambda{\rm CDM} + \varepsilon$ model correspond to net brightening of the source, and therefore cannot be explained by ordinary positive-opacity photon loss alone~\cite{Bassett:2003vu, Bassett:2003zw}. Photon--chameleon scenarios can produce effective brightening under specific assumptions about supernova-explosion conversions and propagation~\cite{Burrage:2007ew}. Nonetheless, the magnitude of this effect depends on poorly constrained scalar production in the supernova and on astrophysical magnetic fields, while subsequent astrophysical and laboratory searches~\cite{Steffen:2010ze} substantially constrain the relevant photon-chameleon coupling employed in Ref.~\cite{Burrage:2007ew}. More general photon conversion, source evolution, or calibration mechanisms require their own propagation or luminosity model, whose study lies outside the scope of this work. CMB-temperature measurements also constrain photon-number non-conservation~\cite{Ruchika:2025sbb}, but translating those limits into distance-duality constraints assumes adiabatic and approximately achromatic photon dimming. Under these assumptions, large violations capable of resolving the calibration discrepancy are strongly constrained, whereas frequency-dependent conversion mechanisms require a separate assessment~\cite{Tiwari:2026pzk}. We note that these bounds only exclude these particular mechanisms but not every possible physical mechanism of an effective CDDR redshift dependence.

\subsection{Effective gravitational coupling}

A change in the effective gravitational coupling $G(z)$ is a possible source-evolution hypothesis~\cite{Amendola:1999vu}, but translating the tilt $\varepsilon$ into a $G(z)$ expression is model dependent. In the absence of a screening mechanism, the effective gravitational constant influences the supernova luminosity $L$, producing a scaling of the form $L/L_0 = {(G(z)/G_0)}^p$ where the subscript zero denotes its local value. This produces a scaling in the absolute magnitude $M_{\rm B}$ of the form~\cite{Riazuelo:2001mg}
\begin{equation}
    M_{\rm B} =
    M_{0{\rm B}} - \frac{5}{2} \log_{10}{(G(z)/G_0)}^p.
\end{equation}
From the SN luminosity distance relation, we can match the redshift-dependent part of $\eta(z)$ with $G(z)$, yielding
\begin{equation}
    G(z)/G_0 =
    {(1+z)}^{-2\varepsilon/p}
    ,\qquad
    (\dot{G}/G) \vert_0 =
    2\varepsilon H_0 / p.
    \label{eq:G_evolution}
\end{equation}

To assess whether $G(z)$ evolution could account for the effective tilt, $\varepsilon$, we use the unscreened relation~\eqref{eq:G_evolution} under the standard Chandrasekhar-mass scaling $p \!=\! -3/2$~\cite{Riazuelo:2001mg, Gaztanaga:2001fh}%
\footnote{We note that this scaling is not a prediction for the fully standardized luminosity. In fact, explosion physics and light-curve standardization can change the magnitude and even the sign of the response~\cite{Wright:2017rsu,Mould:2014iga}.}
and the common tilt $\varepsilon \simeq -0.03$ found in the $\Lambda{\rm CDM} + \varepsilon$ model. Then, for a standard $H_0 \simeq 70\, {\rm km}\, {\rm s}^{-1}\, {\rm Mpc}^{-1}$, the present-day time-evolution of the gravitational coupling $(\dot{G}/G)\vert_0$ would be positive and of order $\sim\!3 \times 10^{-12}\, {\rm yr}^{-1}$. However, lunar-laser ranging constrains such variations much more tightly~\cite{Hofmann:2018myc,Biskupek:2020fem} (at the order of $10^{-13}$--$10^{-14}\, {\rm yr}^{-1}$), challenging the particular unscreened scenario considered. This claim is consistent with earlier findings from Ref.~\cite{Sapone:2020wwz}. However, note that these constraints do not directly exclude a screened coupling, an environmental response, or a transition that has ended before the local measurement epoch~\cite{Desmond:2019ygn, Marra:2021fvf}. This would require a self-consistent model for the background, local gravity, stellar luminosities, and the relevant calibrators before confronting the model with all other data.

\section{\label{sec:conclusions}Conclusions}

In this work we investigated how the preference for a CPL dynamical dark energy ($w_0w_a{\rm CDM}$) over the standard $\Lambda{\rm CDM}$ can be affected by considering violations of the cosmic distance duality relation (CDDR). By assuming a simple power-law parametrization $\eta(z) \equiv D_{\rm L}/[{(1+z)}^2 D_{\rm A}(z)] = \alpha{(1+z)}^\varepsilon$, we analyzed the sensitivity of the dynamical dark energy preference signal under calibration-free DESI DR2 BAO data and four SN compilations (Pantheon$+$, DES-Dovekie, Union3, and Union3.1). We excluded any early-time prior coming from the CMB in this main analysis, thereby leaving $\alpha$ unconstrained.

When allowing for deviations in the CDDR relation for both $\Lambda{\rm CDM}$ and $w_0w_a{\rm CDM}$, the CPL preference with respect to the zero-tilt baseline, $\varepsilon = 0$, is reduced from $1.8 \sigma$ to $1.0\sigma$ for Pantheon$+$, $2.2\sigma$ to $1.5\sigma$ for DES-Dovekie, $2.7\sigma$ to $1.9\sigma$ for Union3, and $2.0\sigma$ to $1.8\sigma$ for Union3.1. The negative individual values of $\varepsilon$ found in the $\Lambda{\rm CDM} + \varepsilon$ setting are weak and background dependent, and the sign of $\varepsilon$ changes when the dark-energy equation of state parameters $(w_0, w_a)$ of the CPL parametrization are freed, mainly because of the correlation between $\varepsilon$ and $w_0$. We quantified that these improvements remained stable under alternative $\eta(z)$ parametrizations, and were capable of producing a larger reduction in the preference for $w_0w_a{\rm CDM}$ than a free spatial curvature $\Omega_k$ affecting both BAO and SN distances, rendering the power-law extension as a compact way to test specific CDDR violations. The only exception is Union3.1, for which any one-parameter extension performs equally well in slightly reducing the preference for dynamical dark energy. In addition, we found that deletion of low-redshift tracers in BAO consistently increased the reduction in the dynamical dark energy signal (see Refs.~\cite{Colgain:2024xqj, Colgain:2025nzf, Colgain:2026ryf} for further implications of these tracers on $\Omega_{\rm m}$ constraints and dynamical dark energy), while higher-redshift ones maintained a similar reduction (ELG2 and QSO) or decreased it (LRG$3+$ELG$1$ and Ly$\alpha$) compared to the full BAO analysis. Moreover, removal of low-redshift samples in the uncompressed SN samples did not affect the conclusions for Pantheon$+$, but greatly altered the significance (before and after freeing $\varepsilon$) for the DES-Dovekie sample.

As a complementary analysis, we employed illustrative calibrations for the sound horizon $r_{\rm d} = 147.09 \pm 0.26\, {\rm Mpc}$ coming from Planck and for the absolute magnitude $M_{\rm B} = -19.253 \pm 0.027$ coming from SH0ES.\@ These, together with BAO$+$Pantheon$+$, provided a constraint on the normalization of $\eta(z)$ of $\alpha \simeq 0.942 \pm 0.014$, quantifying the calibration mismatch conditional on the adopted background and priors. We also included cosmic chronometers (CC) in the analysis, though BAO$+$CC (which are blind to $\varepsilon$) in the absence of external calibrations yielded $r_{\rm d} = 146.6_{-6.6}^{+7.4}\, {\rm Mpc}$ for $\Lambda{\rm CDM}$, too weak to distinguish a sound-horizon shift from a late-time calibration shift. Neither result supports assigning the discrepancy to a unique physical sector. Ordinary positive-opacity photon loss has the wrong sign to explain the negative central tilts $\varepsilon$ found for $\Lambda{\rm CDM} + \varepsilon$, and local gravity tests strongly constrain the illustrative unscreened varying-$G$ interpretation. In addition, CMB-temperature limits further restrict adiabatic, approximately achromatic photon-dimming scenarios. In any case, these statements are conditional on the physical assumptions and do not constitute a general exclusion of source evolution or photon-conversion models.

We acknowledge several limitations of our study that are central to the interpretation of these results. First, the fits performed are for already released, standardized supernova likelihoods, and do not re-estimate selection effects or re-train light-curve and population models when a tilt $\varepsilon$ is introduced. Second, and in a similar vein, the Union3 and Union3.1 samples compress the underlying supernovae magnitudes before accounting for CDDR violations. Their constraints therefore do not guarantee exact equivalence to a full hierarchical refit with the new degree of freedom $\varepsilon$. Third, the four SN samples considered are statistically dependent: DES-Dovekie shares low-redshift supernovae with the Pantheon$+$ sample, whereas the Union releases constitute an alternative compression of an almost overlapping SN catalog with Pantheon$+$. Moreover, the changes from Union3 to Union3.1 include host-galaxy information and population-standardization updates~\cite{Hoyt:2026fve,Rubin:2026qdt}, and they should not be attributed to a single calibration modification as the one that would arise from CDDR violation. Therefore, the overlapping results for the $\varepsilon$ tilt from all samples cannot establish a universal effect or a signal that we can combine into a single overall constraint.

Testing the tilt in full supernova likelihood analyses by re-estimating selection effects or re-training light-curves, and jointly analyzing independent expansion and distance probes are possible future steps to assess the physical significance of CDDR.\@ A reanalysis of the presented results under the new recalibration and combined samples presented by the Unite SN compilation~\cite{Camilleri:2026rer,Lee:2026dum}, when publicly available, would also constitute a natural next step to provide updated constraints.

\begin{acknowledgments}
We are grateful to Guillaume Poggiaspalla and the DIPC Supercomputing Center for providing access to computational resources. M.~A.~acknowledges support from the Basque Government Grant No.~PRE\_2025\_2\_0301 and the Egonlabur mobility program No.~EP\_2026\_1\_0083. This paper is based upon work from COST Action CA21136 \textit{Addressing observational tensions in cosmology with systematics and fundamental physics} (CosmoVerse), supported by COST (European Cooperation in Science and Technology). M.~A.~wishes to thank A.~Nasoni for carefully reading our manuscript and for her valuable feedback.

The authors acknowledge the use of AI assistance to optimize numerical codes, refine figures and selected portions of the text for clarity, and identify and correct typographical errors. All content was subsequently reviewed and verified by the authors, who take full responsibility.
\end{acknowledgments}

\section*{Data Availability}
The numerical codes supporting this analysis are openly available~\cite{Artola:2026git}.

\begin{table}[b]
\caption{Priors for the posterior inference. Uniform ranges, denoted by $\mathcal{U}(a,b)$, apply only when the parameter is free; absent parameters are fixed to their model values. The last two rows denote Gaussian calibration priors, used only in Sec.~\ref{sec:calibrated_constraints_on_CDDR}.}\label{tab:priors}
\begin{ruledtabular}
\begin{tabular}{lll}
Parameter & Prior & Application \\
\hline
$\Omega_{\rm m}$ & $\mathcal U(0.1,0.5)$ & All backgrounds \\
$w_0$ & $\mathcal U(-3,0.5)$ & $w_0w_a{\rm CDM}$ \\
$w_a$ & $\mathcal U(-4,3)$ & $w_0w_a{\rm CDM}$ \\
$\varepsilon$ & $\mathcal U(-0.4,0.4)$ & Tilt free \\
$H_0 r_{\rm d}\, ({\rm km}\, {\rm s}^{-1})$ & $\mathcal U(c/100,c/10)$ & BAO without CC \\
$\nu\ ({\rm mag})$ & $\mathcal U(10,60)$ & SN intercept \\
$H_0\ ({\rm km}\, {\rm s}^{-1}\, {\rm Mpc}^{-1})$ & $\mathcal U(20,120)$ & BAO with CC \\
$r_{\rm d}\ ({\rm Mpc})$ & $\mathcal U(50,300)$ & With CC \\
$r_{\rm d}\ ({\rm Mpc})$ & $147.09 \pm 0.26$ & Calibration only \\
$M_{\rm B}\ ({\rm mag})$ & $-19.253 \pm 0.027$ & Calibration only \\
\end{tabular}
\end{ruledtabular}
\end{table}

\begin{table*}[t]
\caption{Marginalized posterior medians and 16th/84th-percentile errors for DESI DR2 BAO plus each SN compilation. The labels in the first column are understood to be combined with the BAO data. The $\chi_{\min}^2$ column contains separately optimized likelihood minima.}\label{tab:posterior}
\footnotesize
\begin{ruledtabular}
\begin{tabular}{llrcccc}
Sample & Model & $\chi^2_{\min}$ & $\Omega_{\rm m}$ & $w_0$ & $w_a$ & $\varepsilon$ \\
\hline\rule{0pt}{1.4em}%
Pantheon$+$ & $\Lambda{\rm CDM}$ & $1399.80$ & $0.3046^{+0.0080}_{-0.0079}$ & $-1$ & $0$ & $0$ \\[0.37em]
 & $w_0w_a{\rm CDM}$ & $1394.74$ & $0.303^{+0.016}_{-0.028}$ & $-0.885^{+0.062}_{-0.059}$ & $-0.16^{+0.55}_{-0.45}$ & $0$ \\[0.37em]
 & $\Lambda{\rm CDM} + \varepsilon$ & $1396.74$ & $0.2976^{+0.0089}_{-0.0085}$ & $-1$ & $0$ & $-0.029^{+0.016}_{-0.017}$ \\[0.37em]
 & $w_0w_a{\rm CDM} + \varepsilon$ & $1394.56$ & $0.307^{+0.020}_{-0.031}$ & $-0.85^{+0.12}_{-0.11}$ & $-0.26^{+0.64}_{-0.54}$ & $0.010^{+0.035}_{-0.035}$ \\[0.37em]
\hline\rule{0pt}{1.4em}%
DES-Dovekie & $\Lambda{\rm CDM}$ & $1645.28$ & $0.3062^{+0.0078}_{-0.0073}$ & $-1$ & $0$ & $0$ \\[0.37em]
 & $w_0w_a{\rm CDM}$ & $1638.12$ & $0.313^{+0.013}_{-0.018}$ & $-0.845^{+0.072}_{-0.065}$ & $-0.51^{+0.50}_{-0.46}$ & $0$ \\[0.37em]
 & $\Lambda{\rm CDM} + \varepsilon$ & $1641.48$ & $0.2974^{+0.0085}_{-0.0086}$ & $-1$ & $0$ & $-0.029^{+0.015}_{-0.015}$ \\[0.37em]
 & $w_0w_a{\rm CDM} + \varepsilon$ & $1637.40$ & $0.323^{+0.018}_{-0.021}$ & $-0.76^{+0.13}_{-0.12}$ & $-0.74^{+0.54}_{-0.52}$ & $0.025^{+0.032}_{-0.033}$ \\[0.37em]
\hline\rule{0pt}{1.4em}%
Union3 & $\Lambda{\rm CDM}$ & $38.81$ & $0.3038^{+0.0086}_{-0.0081}$ & $-1$ & $0$ & $0$ \\[0.37em]
 & $w_0w_a{\rm CDM}$ & $28.78$ & $0.330^{+0.015}_{-0.018}$ & $-0.70^{+0.12}_{-0.11}$ & $-1.00^{+0.57}_{-0.56}$ & $0$ \\[0.37em]
 & $\Lambda{\rm CDM} + \varepsilon$ & $34.18$ & $0.2977^{+0.0085}_{-0.0086}$ & $-1$ & $0$ & $-0.048^{+0.023}_{-0.022}$ \\[0.37em]
 & $w_0w_a{\rm CDM} + \varepsilon$ & $28.43$ & $0.338^{+0.022}_{-0.024}$ & $-0.63^{+0.17}_{-0.17}$ & $-1.20^{+0.68}_{-0.68}$ & $0.020^{+0.037}_{-0.037}$ \\[0.37em]
\hline\rule{0pt}{1.4em}%
Union3.1 & $\Lambda{\rm CDM}$ & $41.09$ & $0.3021^{+0.0081}_{-0.0081}$ & $-1$ & $0$ & $0$ \\[0.37em]
 & $w_0w_a{\rm CDM}$ & $34.68$ & $0.322^{+0.015}_{-0.020}$ & $-0.77^{+0.11}_{-0.10}$ & $-0.77^{+0.59}_{-0.55}$ & $0$ \\[0.37em]
 & $\Lambda{\rm CDM} + \varepsilon$ & $38.93$ & $0.2976^{+0.0086}_{-0.0085}$ & $-1$ & $0$ & $-0.031^{+0.021}_{-0.021}$ \\[0.37em]
 & $w_0w_a{\rm CDM} + \varepsilon$ & $33.67$ & $0.335^{+0.021}_{-0.023}$ & $-0.66^{+0.17}_{-0.16}$ & $-1.12^{+0.68}_{-0.64}$ & $0.035^{+0.038}_{-0.038}$ \\[0.37em]
\end{tabular}
\end{ruledtabular}
\end{table*}

\appendix
\section{\label{app:constraints} Priors, and one- and two-dimensional constraints}

\begin{figure*}[t]
    \includegraphics{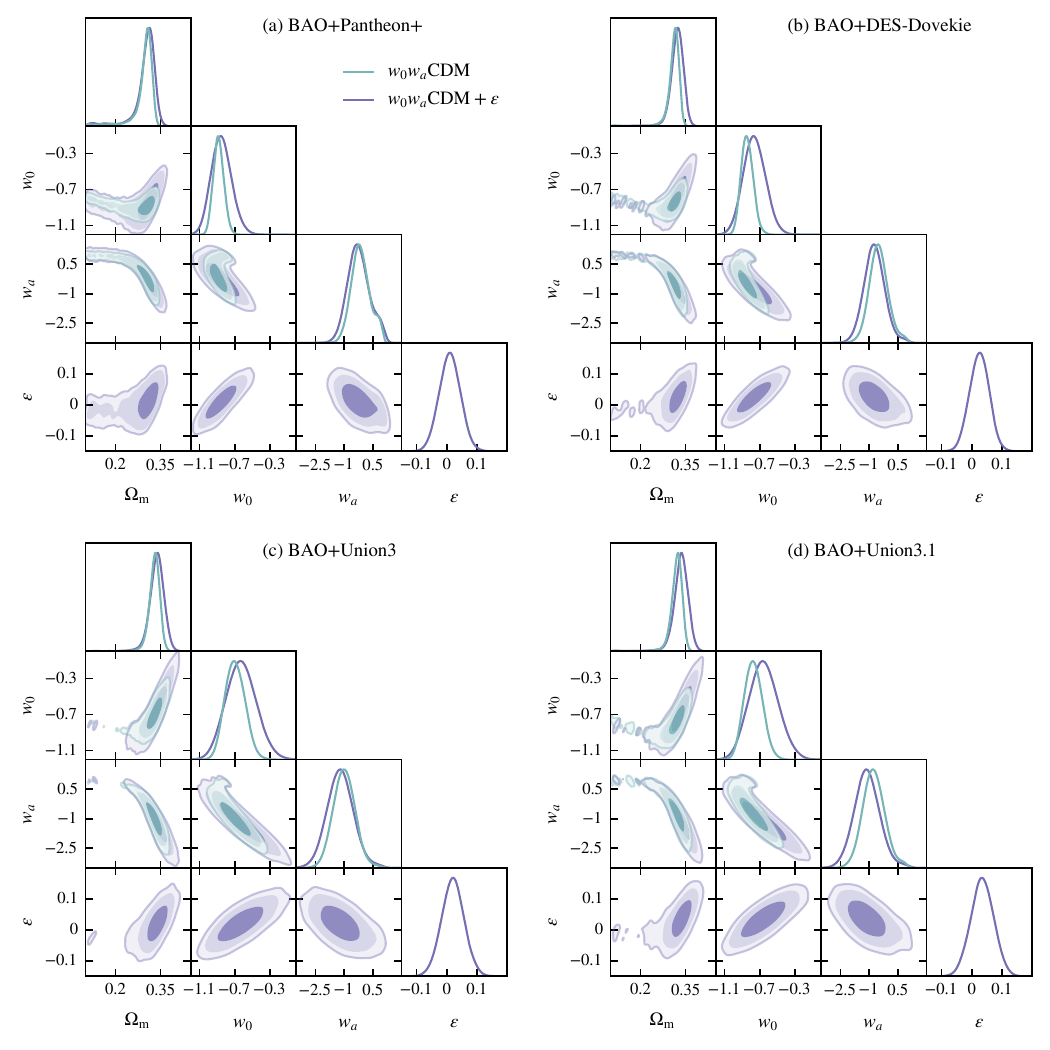}
    \caption{One- and two-dimensional marginalized posterior distributions for the $w_0w_a{\rm CDM}$ and $w_0w_a{\rm CDM} + \varepsilon$ models when combining DESI DR2 BAO with each SN catalog. The legend entry in the (a) panel is shared across panels (b)--(d). $68$, $95$ and $99\%$ probability contours are reported; the latter allow for a better visualization of the degeneracies in this extended model. The contours show non-Gaussian degeneracies within the adopted prior domain, especially in the $\Omega_{\rm m}$ one- and two-dimensional marginalized posteriors.}\label{fig:2x2_posterior}
\end{figure*}

Table~\ref{tab:priors} provides the uniform priors we employed in our sampling with \textsc{Nautilus}. We have checked that increasing the prior ranges does not affect the constraints and conclusions. We also collect the results from Sec.~\ref{sec:results} here. Table~\ref{tab:posterior} contains the marginalized posterior medians, as well as the $16^{\rm th}$ and $84^{\rm th}$ percentile asymmetric errors for the relevant cosmological parameters $\boldsymbol{\theta} = (\Omega_{\rm m}, w_0, w_a, \varepsilon)$ for DESI DR2 BAO combined with each SN sample. Fig.~\ref{fig:2x2_posterior} complements the table with one- and two-dimensional marginalized posteriors for $w_0w_a{\rm CDM}$ and $w_0w_a{\rm CDM} + \varepsilon$, with contours enclosing $68\%$, $95\%$, and $99\%$ probability regions. For Pantheon$+$ and DES-Dovekie, the two-dimensional posteriors including $\Omega_{\rm m}$ present strong non-Gaussianity compared with the constraints coming from the Union catalogs.

\bibliography{Bibliography}
\bibliographystyle{apsrev4-2-titles}
\end{document}